\documentclass[letterpaper,english,reprint, aps]{revtex4-1}
\usepackage[T1]{fontenc}
\usepackage[latin9]{inputenc}
\usepackage{amsmath}
\usepackage{xcolor}
\usepackage{graphicx}
\usepackage{amssymb}

\makeatletter

\pdfpageheight\paperheight
\pdfpagewidth\paperwidth

\makeatother

\usepackage{babel}
\begin{document}
\title{Pulse-timing symmetry breaking in an excitable optical system with delay}

\author{Soizic Terrien $^1$}
\email{s.terrien@auckland.ac.nz}

\author{Venkata A. Pammi $^2$}

\author{Bernd Krauskopf $^1$}

\author{Neil G.R. Broderick $^1$}

\author{Sylvain Barbay $^2$}

\affiliation{$^1$The Dodd-Walls Centre for Photonic and Quantum Technologies, The University of Auckland, New Zealand~\\
$^2$Universit\'e Paris-Saclay, CNRS, Centre de Nanosciences et de Nanotechnologies, Palaiseau, France.}
\date{\today}
\begin{abstract}
Excitable systems with delayed feedback are important in areas from biology to neuroscience and optics. They sustain multistable pulsing regimes with different number of equidistant pulses in the feedback loop. Experimentally and theoretically, we report on the pulse-timing symmetry breaking of these regimes in an optical system. A bifurcation analysis unveils that this originates in a resonance phenomenon and that symmetry-broken states are stable in large regions of the parameter space. These results have impact in photonics for e.g. optical computing and versatile sources of optical pulses.
\begin{description}
\item [{PACS~numbers}] May be entered using the environment \textsf{PACS~numbers}.
\end{description}
\end{abstract}
\maketitle

\paragraph*{Introduction}

Time periodic regular pulsing regimes can emerge in many dissipative physical systems with delayed feedback \citep{PyragasPLA92,Erneux09}. This phenomenon is encountered in various fields, from neurosciences \citep{LongtinMB88,Campbell07} to optics and opto-electronics \citep{IkedaPRL82,IllingPD05,KouomouPRL05,SorianoRMP13}, ecology \citep{HutchinsonANAS48} and chemistry \citep{EpsteinJCP91,RousselJPC96}. Typically, these systems are multistable, with several coexisting regular periodic regimes \citep{FossPRL96,HizanidisIJBC08}. Multistability has been shown to be of particular interest for all-optical processing capabilities, e.g. associative memories \citep{prucnal2016recent,nahmias2013leaky,TerrienPRR20}.

Here we consider an excitable optical system with delayed feedback, namely a micropillar laser with integrated saturable absorber and delayed optical feedback. In the excitable regime and for a sufficiently large delay time, the system  regenerates its own output at regular time intervals \citep{GarbinNC15,RomeiraNSR16,TerrienPRA17,TerrienOL18}: if a short duration perturbation with sufficiently large amplitude is sent as an input, the system emits a light pulse which is re-injected by the feedback loop after a time delay $\tau$. If the losses in the feedback loop are sufficiently low, the re-injected pulse is regenerated in the excitable medium. As the process repeats, this results in a periodic pulsing regime with period $T$ slightly larger than $\tau$ due to the finite response time of the excitable medium. 

When several perturbations are sent sequentially, the timing structure of the regenerated pulses persist in the short term. This can lead to interesting applications such as optical buffer memories \citep{GarbinNC15,RomeiraNSR16,TerrienPRA17,TerrienOL18}. However, in the long term, the system must converge to one of the coexisting stable pulsing regimes. These regimes generally have period $T_{n}$ close to sub-multiples of $\tau$ and, as such, consist of roughly $n$ \textit{equidistant} pulses in the feedback loop \citep{TerrienPRR20}. 

In this Letter we report on a different stable asymptotic behavior, where the long-term dynamics consists of non-equidistant pulses in the feedback loop. This is observed experimentally and in a model, namely a system of three coupled delay-differential equations (DDEs). A bifurcation analysis unveils that the observed non-equidistant pulsing dynamics results from destabilising bifurcations of the equidistant pulsing solutions; these occur when the delay $\tau$ is increased, provided that the recombination rate of carriers in the gain medium is faster than the one of the saturable absorber. We show that the emergence of stable pulsing patterns with $n$ non-equidistant pulses is generic in the system and arises from symmetry breaking due to locked dynamics on invariant tori. Two stable non-equidistant pulses in the feedback loop emerge from a period-doubling bifurcation of the equidistant two-pulse solution, which is close to a $1\!:\!2$ strong resonance. Because of the amplitude-timing coupling in the excitable system \citep{SelmiPRE16,ErneuxPRE18}, the relative timings of the pulses are very strongly affected. This is responsible for the observed immediate symmetry breaking of pulse timings. Pulsing patterns with $n$ equidistant pulses per feedback loop with $n\geq3$ destabilize in torus bifurcations that are close to $1\!:\!n$ resonance. This results in large resonance tongues, \emph{i.e.} regions of the parameter space where the dynamics on the torus is locked. Due to the amplitude-timing coupling, these locked $1\!:\!n$ periodic orbits correspond, here, to higher-order symmetry-broken, that is, non-equidistant pulsing regimes. We show that non-equidistant and equidistant pulsing regimes can coexist, thus leading to a much increased level of multistability of pulsing patterns.


\paragraph*{Model equations}

We consider the Yamada equations with incoherent delayed feedback \cite{Yamada93, KrauskopfWalker}, a model of semiconductor laser written in the form of three DDEs for the dimensionless gain $G$, absorption $Q$ and intracavity intensity $I$:
\begin{equation}
\begin{split}
\dot G&= \gamma_G (A-G-GI);\\
\dot Q&= \gamma_Q(B-Q-sQI);\\
\dot I&=(G-Q-1)I + \kappa I(t-\tau).
\label{eq:yamada}
\end{split}
\end{equation}
The time variables are rescaled with respect to the cavity photon lifetime \cite{BarbayOL11}. Here, $A$ is the pump parameter, $B$ is the non-saturable absorption, $s$ is the scaled saturation parameter, $\gamma_G$ and $\gamma_Q$ are the recombination rates of the carriers in the gain and absorber media, respectively, and $\kappa$ and $\tau$ are the feedback strength and delay, respectively. Unless stated otherwise, we consider the following parameters values: $A=2$, $B=2$, $\gamma_G=0.01$, $\gamma_Q=0.055$, $s=10$, and $\kappa=0.2$. The delay time $\tau$ is considered as a bifurcation parameter. This model has been shown to produce rich and complex dynamics \cite{KrauskopfWalker,TerrienSIADS17}, and to describe accurately the dynamics of an excitable micropillar laser with integrated saturable absorber subject to delayed optical feedback \cite{TerrienOL18,TerrienPRR20}.

\begin{figure}[t!]
\includegraphics[width=\linewidth]{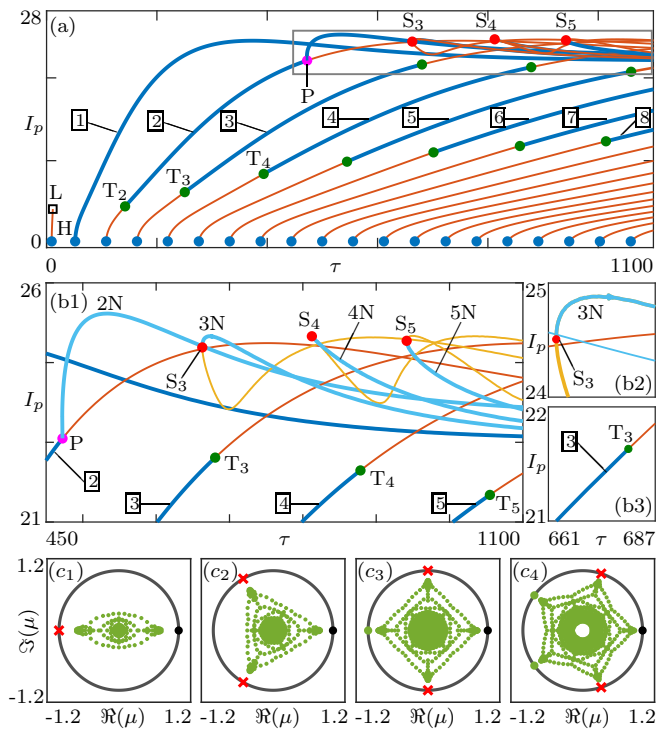}
\caption{(a) Bifurcation diagram of \eqref{eq:yamada}, showing the pulse intensity $I_p$ with respect to $\tau$, with the number of pulses per feedback loop along each stable periodic solution branch. (b1) Enlargement of the framed area in (a), with further enlargments around point T$_3$ (b2) and S$_3$ (b3). Stable equidistant (E) and non-equidistant (N) pulse solutions are represented in dark and light blue, respectively, and unstable E and N solutions in dark and light orange, respectively. The dots indicate Hopf (H), torus (T), period doubling (P), saddle-node (S) and homoclinic (L) bifurcations. (c1)--(c4) Floquet multipliers at points P, T$_3$, T$_4$ and T$_5$, with critical multipliers highlighted in red.}
\label{fig:bifdiag1D}
\end{figure}

\paragraph*{Bifurcation analysis}

Figure \ref{fig:bifdiag1D}(a) shows the one-parameter bifurcation diagram of system \eqref{eq:yamada} in the delay time $\tau$, where solutions are represented by their maximum value $I_p$ of intensity $I$. When $\tau$ is increased from zero, successive Hopf bifurcations (H) are encountered, from which several branches of coexisting periodic solutions emerge. Far from the Hopf bifurcations, these solutions correspond to the periodic emission of short light pulses, in between which the intensity is practically zero, with periods close to submultiples of the delay \cite{TerrienSIADS17}. Each of these solutions thus corresponds to a fixed number of equidistant pulses in the feedback loop, as indicated by the numbering in Figure \ref{fig:bifdiag1D}(a). The fundamental solution with one pulse per feedback loop appears at $\tau=51.7$ and is stable for any larger value of $\tau$. On the other hand, all the $n$-pulses solutions with $n \geq 2$ emerge unstably from a Hopf bifurcation, subsequently stabilize in a torus bifurcation (T) when $\tau$ is increased, and finally destabilize through a second bifurcation. All these solutions coexist with the zero-intensity equilibrium solution (\textit{i.e.} the non-lasing solution), which is stable over the entire range of $\tau$ in Figure \ref{fig:bifdiag1D}(a).

Figure \ref{fig:bifdiag1D}(b) present enlargements of panel (a) near the second (destabilizing) bifurcations of the pulsing regimes with two to five equidistant pulses (points P, T$_3$, T$_4$ and T$_5$, respectively), and panels (c) show the Floquet multipliers of the pulsing solutions at these points. The loss of stability of the two-pulse solution occurs at point P through a period-doubling bifurcation, with one Floquet multiplier crossing the unit circle at -1 [panel (c1)]. This bifurcation is close to a $1\!:\!2$ resonance point, where two Floquet multipliers are equal to $-1$ \cite{Kuznetsov2013elements}; this has been checked by varying slightly the value of the feedback strength $\kappa$ (not shown). The three-, four- and five-pulse solutions destabilize at points T$_3$, T$_4$ and T$_5$, respectively, at torus bifurcations, where two complex conjugate Floquet multipliers cross the unit circle. Panels (c2)--(c4) show that these critical Floquet multipliers are extremely close to (but not quite equal to) $e^{\pm i 2\pi /n}$, which means that the torus bifurcations that destabilize the equidistant solutions are very close to $1\!:\!n$ resonance points with $n=$3, 4 and 5, respectively. As a result, the bifurcating stable multifrequency dynamics on the invariant torus is almost immediately $1\!:\!n$ locked; this happens at saddle-node bifurcations S of periodic solutions. 

\begin{figure}[t!]
\includegraphics[trim= 0cm 0cm 0cm 0cm, clip=true, width=\linewidth]{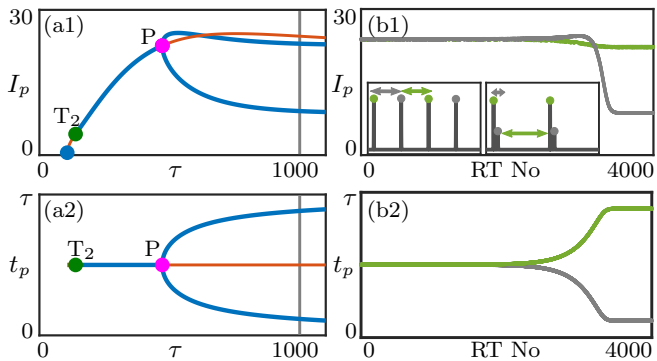}
\caption{(a) Maximum $I_p$ of pulse intensity (a1) and relative interpule timings $t_p$ (a2) along the branches of two equidistant and non-equidistant pulses, with respect to $\tau$. Stable and unstable solutions are represented in blue and red, respectively. (b) Simulation of \eqref{eq:yamada} for $\tau=1000$ (grey lines in panel (a)) with initial condition very near the (unstable) two-pulse solution, showing the long-term evolution of $I_p$ (b1) and of $t_p$ (b2). The subpanels in (b1) show the intensity time series during the two first and two last roundtrips through the feedback loop; the dots and arrows indicate the amplitudes and relative timings as represented in (b1) and (b2), respectively.}
\label{fig:CVnum}
\end{figure}

As Figure \ref{fig:bifdiag1D}(b1) shows, these two bifurcation mechanisms lead to the emergence of additional periodic solutions. A branch of stable (period-doubled) periodic solutions emerges from the period-doubling bifurcation of the two-pulse solution. This solution of period close to $\tau$ corresponds to two non-equidistant pulses per feedback loop, and thus appears as a pulse-timing symmetry broken state. This is illustrated in Figure \ref{fig:CVnum}(a), which shows the evolution of the amplitudes and relative timings of pulses with respect to $\tau$, along the branches of the two equidistant and non-equidistant solutions. After the period doubling bifurcation at $\tau=472$, one observes both a splitting of the pulse amplitudes in panel (a1) (as expected), but also a strong splitting or symmetry breaking of the relative pulse timings [panel (a2)], which is due to the strong time-amplitude coupling \cite{YanchukPRL19,TerrienPRR20}. 

We use numerical simulations to further assess how the regime with two non-equidistant pulses is accessed. Figure \ref{fig:CVnum}(b) shows, for $\tau=1000$, the long-term dynamics of  system \eqref{eq:yamada} when it is initially on the (unstable) equidistant two-pulse solution and subsequently slightly perturbed by increasing the gain variable $G$. The system slowly converges to one of the two possible non-equidistant stable pulsing patterns: the first pulse timing interval decreases [panel b2] and the second pulse (highlighted in gray) converges towards a low amplitude state [panel b1]. When a different initial perturbation is applied by depleting $G$ slightly (not shown here), the phase-shifted, symmetric version of this solution is obtained, with the first (green) and second (gray) pulses converging to the low-amplitude and high-amplitude state, respectively. Although this leads seemingly to the same long-term dynamics, both of these different states occur, one being a phase-shifted version of the other. We also point out that the convergence is very slow and occurs over several thousands of delay times, showing that the stable non-equidistant solutions are only weakly attracting.

The bifurcation mechanism leading to the emergence of non-equidistant pulsing regimes with more than three pulses is slightly different. As shown in Figure \ref{fig:bifdiag1D}(b1), a pair of (stable/unstable) periodic solutions emerges from a saddle-node bifurcation, for example at $\tau=663$ for $n=3$. This bifurcation forms the boundary of the $1\!:\!n$ resonance tongue associated with the destabilizing torus bifurcation of the $n$-pulse solution. The emerging periodic solutions have a period close to $\tau$, compared to the period close to $\tau/n$ of the $n$-pulse solution undergoing the torus bifurcation. Here, the stable $1\!:\!n$ locked periodic solution corresponds to a pulsing regime with $n$ non-equidistant pulses of different amplitude in the feedback loop. As such, the $1\!:\!n$ resonance tongues are identified here as the stability regions of non-equidistant pulsing solutions. Their emergence leads to a rapidly increasing level of multistability. 

\begin{figure*}[ht]
\includegraphics[width=\textwidth]{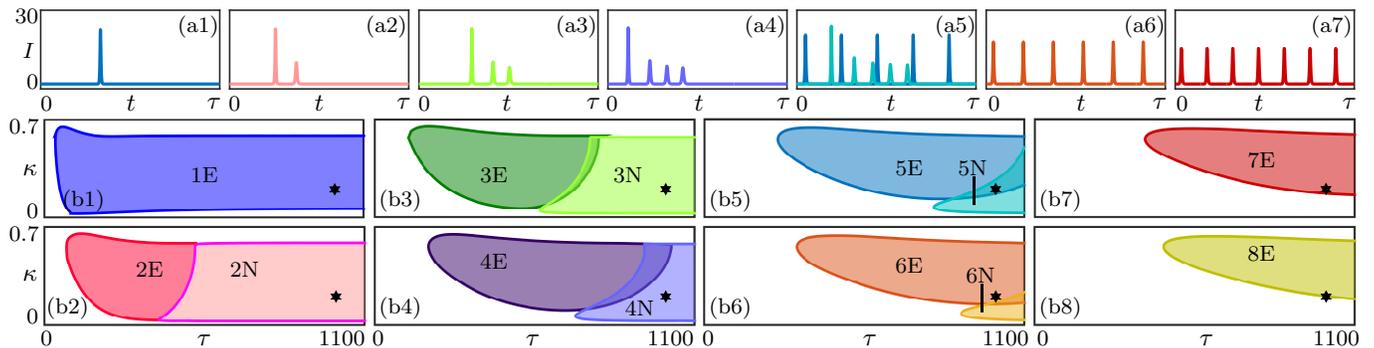}
\caption{ (a) Intensity profiles of stable periodic solutions of \eqref{eq:yamada}, for $\tau=1000$. (b) Regions of stability, in the $(\tau,\kappa)$-plane of feedback parameters, of the families of equidistant (E) and non-equidistant (N) periodic solutions with 1 to 8 pulses per feedback loop. The number of pulses is indicated in the colored regions, and the star indicates the parameter point $(\tau,\kappa)=(1000,0.2)$ of the time series in panels (a).}
\label{fig:intensity}
\end{figure*}

Figure \ref{fig:intensity}(a) shows the intensity profiles of the coexisting stable periodic solutions for $\tau=1000$. Here the non equidistant two-, three-, four- and five-pulse solutions [panels (a2)--(a5)] coexist with the stable one-pulse solution [panel (a1)], but also with the stable solutions with 5, 6 and 7 equidistant pulses in the feedback loop [panels (a5)--(a7)]. Overall, when $\tau$ is increased, more and more of the equidistant pulsing regimes become unstable, while more and more stable periodic solutions with non-equidistant pulses in the feedback loop appear. Typically, for sufficiently large $\tau$, solutions with lower numbers of (at least two) non-equidistant pulses coexist with solutions with larger numbers of equidistant pulses. In Figure \ref{fig:intensity}, all the periodic solutions with 1 to 7 pulses per feedback loop coexist, but the ones with two to five pulses already underwent the resonance tongue transition and, thus, correspond to non-equidistant pulsing patterns.

Figure \ref{fig:intensity}(b) presents the regions of stability in the $(\tau,\kappa)$-plane of feedback parameters of the different pulsing regimes with one up to eight (equidistant and non equidistant) pulses per feedback loop. Here the regions of stability of the non-equidistant pulsing solutions are resonance tongues bounded by saddle-node bifurcations. The respective stability regions of both types of solutions extend over large areas of the $(\tau,\kappa)$ parameter plane. Moreover, they show a large degree of overlap, which is why we show them in individual panels (b1)--(b8) for one up to eight pulses per feedback loop. This represents the high degree of multistability between all the different solutions represented in panels (a); indeed, the long-term convergence to one or the other pulsing solution depends on the chosen initial conditions. Interestingly, for $n\geq3$ both the solutions with $n$ equidistant pulses and with $n$ non-equidistant pulses may coexist and be stable for the same parameter values. 
\noindent
As shown in Figure 1(b1), this results from the fact that the 1:$n$ resonance tongues are entered (at points S) slightly before the $n$-pulse solutions  destabilize at torus bifurcation points T. Hence, in these ranges of $\tau$, depending on the initial condition,
one observes a pattern with $n$ either equidistant or non-equidistant pulses. In particular, in Figure 3(a5) both the solutions with 5 equidistant and 5 non-equidistant pulses coexist for the considered parameters; see also panel (b5).

\color{black}
\paragraph*{Experimental realization}

We compare the results of the bifurcation analysis with experimental measurements of an excitable micropillar laser. It consists of two gain and one saturable absorber (SA) quantum wells \cite{ElsassEPJD10,SelmiPRL14}, 
is optically pumped at $800$ nm and emits light around $980$ nm. Part of the output light is reinjected into the microlaser after a delay $\tau$, through free-space propagation and after reflection on a mirror. The micropillar laser is perturbed by short optical perturbations of 80 ps duration from a mode-locked Ti:Sa laser emitting around 800 nm. In the absence of feedback, the micropillar laser is in the excitable regime \cite{DubbeldamPRE99}, where the steady state intensity $I$ is zero, but a single high-amplitude, short pulse of light can be emitted in response to an external perturbation of sufficient amplitude \cite{IzhikevichBook,BarbayOL11}. In the presence of feedback, an excitable pulse is regenerated when it is reinjected by the delay loop after the delay $\tau$, thus resulting in the regular emission of light pulses at a period close to $\tau$ \cite{TerrienPRA17,TerrienOL18}.

In system \eqref{eq:yamada} the ratio of the recombinations rates $\gamma_G$ and $\gamma_Q$ of the gain and SA media, respectively, play a crucial role in the pulsing dynamics \cite{DubbeldamOC99,SelmiPRE16,otupiri2019yamada}. In particular, the pulse-timing symmetry breaking is observed only for a faster gain recombination, that is, for  $\gamma_{G}>\gamma_{Q}$.  Experimentally, this parameter regime can be accessed by selecting a suitable micropillar laser (from many on the same chip) by taking advantage of the spread of physical parameters in the course of the nanofabrication process.

\begin{figure}[t!]
\includegraphics[trim= 0cm 0cm 0cm 0cm, clip=true,width=\linewidth]{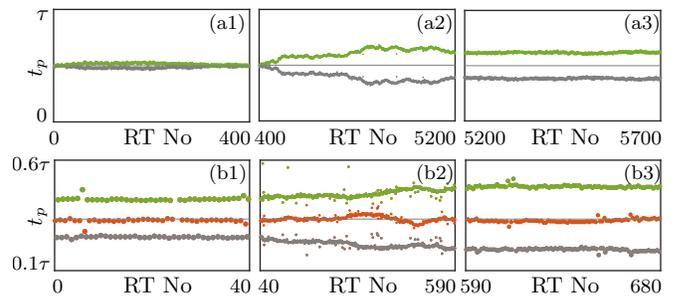}
\caption{Evolution over the roundtrip number of the relative interpulse timing $t p$ of experimental pulse trains following two (a) and three (b) external perturbations, for a feedback delay of $\tau=8.2 ns$: just after the external perturbation (panels 1), during the convergence towards non-equidistant pulsing patterns (panels (2)) and in the long-term (panels 3)).}
\label{fig:expe}
\end{figure}

Figure \ref{fig:expe} shows the evolution of relative pulse timings of experimental pulse trains over several hundreds of roundtrips in the feedback loop, in the same representation as in Figure \ref{fig:CVnum}(b2). The microlaser is started (after suitable external perturbations) with two [panels (a)] and three [panels (b)] almost equidistant pulses in the feedback
loop. We observe that the pulse-timing information is preserved in the short term \cite{TerrienOL18,TerrienPRR20}. On the other hand, in the long term, the system  slowly converges towards non-equidistant pulsing patterns with well-defined and different interpulse relative timings. These interpulse timings then stay very stable over a large number of roundtrip. It was not possible to monitor the amplitude difference in the final state due to the limited signal to noise ratio --- the emitted pulse energy is on the order of only 100 fJ. 
On the other hand, in agreement with Figure \ref{fig:CVnum}, even a small difference in amplitude is associated with a large interpulse interval difference in the non equidistant pulsing regime. Overall, the experimental observations show excellent agreement with the dynamics predicted by the bifurcation analysis of the model.

\paragraph*{Discussion and conclusions}

We demonstrated that an optical excitable system with delayed feedback can sustain stable pulsing patterns with different numbers of non-equidistant pulses in the feedback loop. These arise from stable solutions with $n$ equidistant pulses via torus bifurcations and associated $1\!:\!n$ resonances, which manifest themselves as a swift breaking of the timing-symmetry due to the strong amplitude-timing coupling of the excitable system. We find stable non-equidistant pulsing in large resonance tongues in the parameter space, bounded by saddle-node bifurcations of periodic solutions. As the delay is increased, there is a high and increasing degree of multistability between both symmetric and symmetry-broken pulsing patterns in the feedback loop. Which long-term behavior is observed depends on the initial condition. We demonstrated that non-equidistant pulsing can be observed reliably in an experiment with an excitable micropillar laser.

Our results are reminiscent of the pulsing dynamics of models describing delay-coupled neuron by either two limit cycle oscillators coupled through a time dependent synaptic response \citep{VreeswijkJCN94,BressloffSJAM00} or a pulsing oscillator with delayed self-coupling \citep{KlinshovPRL15,KlinshovPRE15}. 
In our case however, oscillations do not pre-exist and originate from the delayed feedback itself and their period is intimately linked to the delay time. This further illustrates that our results are expected to be generic and to extend beyond optics. Moreover, a mathematical connection between temporal dissipative solitons in spatially extended systems and pulsing regimes in delay systems has been recently suggested \citep{YanchukPRL19}. This raises open questions on possible connections between non-equidistant pulsing regimes and soliton molecules, which are bound states of pulses \citep{Grelu08,KrupaPRL17}.

Beyond their fundamental interest for the nonlinear dynamics of delay systems, our results may contribute to the realization of non-conventional pulsing and reconfigurable optical sources \cite{AadhiOE19}, and to optical computing schemes  \cite{RomeiraC17,Peng19,PammiIJSTQE20,RobertsonIJSTQE20}  that rely on the large phase space available in delay systems \cite{SandeN17}.

\paragraph*{Acknowledgments}

S.B. and V.A.P. acknowledge support from the CNRS Renatech Network of Technology for the nanofabrication of the samples.

\bibliographystyle{apsrev4-1}
\bibliography{timingSSB}

\begin{thebibliography}{46}%
\makeatletter
\providecommand \@ifxundefined [1]{%
 \@ifx{#1\undefined}
}%
\providecommand \@ifnum [1]{%
 \ifnum #1\expandafter \@firstoftwo
 \else \expandafter \@secondoftwo
 \fi
}%
\providecommand \@ifx [1]{%
 \ifx #1\expandafter \@firstoftwo
 \else \expandafter \@secondoftwo
 \fi
}%
\providecommand \natexlab [1]{#1}%
\providecommand \enquote  [1]{``#1''}%
\providecommand \bibnamefont  [1]{#1}%
\providecommand \bibfnamefont [1]{#1}%
\providecommand \citenamefont [1]{#1}%
\providecommand \href@noop [0]{\@secondoftwo}%
\providecommand \href [0]{\begingroup \@sanitize@url \@href}%
\providecommand \@href[1]{\@@startlink{#1}\@@href}%
\providecommand \@@href[1]{\endgroup#1\@@endlink}%
\providecommand \@sanitize@url [0]{\catcode `\\12\catcode `\$12\catcode
  `\&12\catcode `\#12\catcode `\^12\catcode `\_12\catcode `\%12\relax}%
\providecommand \@@startlink[1]{}%
\providecommand \@@endlink[0]{}%
\providecommand \url  [0]{\begingroup\@sanitize@url \@url }%
\providecommand \@url [1]{\endgroup\@href {#1}{\urlprefix }}%
\providecommand \urlprefix  [0]{URL }%
\providecommand \Eprint [0]{\href }%
\providecommand \doibase [0]{http://dx.doi.org/}%
\providecommand \selectlanguage [0]{\@gobble}%
\providecommand \bibinfo  [0]{\@secondoftwo}%
\providecommand \bibfield  [0]{\@secondoftwo}%
\providecommand \translation [1]{[#1]}%
\providecommand \BibitemOpen [0]{}%
\providecommand \bibitemStop [0]{}%
\providecommand \bibitemNoStop [0]{.\EOS\space}%
\providecommand \EOS [0]{\spacefactor3000\relax}%
\providecommand \BibitemShut  [1]{\csname bibitem#1\endcsname}%
\let\auto@bib@innerbib\@empty
\bibitem [{\citenamefont {Pyragas}(1992)}]{PyragasPLA92}%
  \BibitemOpen
  \bibfield  {author} {\bibinfo {author} {\bibfnamefont {K.}~\bibnamefont
  {Pyragas}},\ }\href {\doibase 10.1016/0375-9601(92)90745-8} {\bibfield
  {journal} {\bibinfo  {journal} {Phys. Lett. A}\ }\textbf {\bibinfo {volume}
  {170}},\ \bibinfo {pages} {421} (\bibinfo {year} {1992})}\BibitemShut
  {NoStop}%
\bibitem [{\citenamefont {Erneux}(2009)}]{Erneux09}%
  \BibitemOpen
  \bibfield  {author} {\bibinfo {author} {\bibfnamefont {T.}~\bibnamefont
  {Erneux}},\ }\href {\doibase 10.1007/978-0-387-74372-1} {\emph {\bibinfo
  {title} {Applied Delay Differential Equations}}}\ (\bibinfo  {publisher}
  {Springer New York},\ \bibinfo {year} {2009})\BibitemShut {NoStop}%
\bibitem [{\citenamefont {Longtin}\ and\ \citenamefont
  {Milton}(1988)}]{LongtinMB88}%
  \BibitemOpen
  \bibfield  {author} {\bibinfo {author} {\bibfnamefont {A.}~\bibnamefont
  {Longtin}}\ and\ \bibinfo {author} {\bibfnamefont {J.~G.}\ \bibnamefont
  {Milton}},\ }\href {\doibase 10.1016/0025-5564(88)90064-8} {\bibfield
  {journal} {\bibinfo  {journal} {Math. Biosci.}\ }\textbf {\bibinfo {volume}
  {90}},\ \bibinfo {pages} {183} (\bibinfo {year} {1988})}\BibitemShut
  {NoStop}%
\bibitem [{\citenamefont {Campbell}(2007)}]{Campbell07}%
  \BibitemOpen
  \bibfield  {author} {\bibinfo {author} {\bibfnamefont {S.~A.}\ \bibnamefont
  {Campbell}},\ }in\ \href {\doibase 10.1007/978-3-540-71512-2_2} {\emph
  {\bibinfo {booktitle} {Understanding Complex Systems}}}\ (\bibinfo
  {publisher} {Springer Berlin Heidelberg},\ \bibinfo {year} {2007})\ pp.\
  \bibinfo {pages} {65--90}\BibitemShut {NoStop}%
\bibitem [{\citenamefont {Ikeda}\ and\ \citenamefont
  {Akimoto}(1982)}]{IkedaPRL82}%
  \BibitemOpen
  \bibfield  {author} {\bibinfo {author} {\bibfnamefont {K.}~\bibnamefont
  {Ikeda}}\ and\ \bibinfo {author} {\bibfnamefont {O.}~\bibnamefont
  {Akimoto}},\ }\href {\doibase 10.1103/physrevlett.48.617} {\bibfield
  {journal} {\bibinfo  {journal} {Phys. Rev. Lett.}\ }\textbf {\bibinfo
  {volume} {48}},\ \bibinfo {pages} {617} (\bibinfo {year} {1982})}\BibitemShut
  {NoStop}%
\bibitem [{\citenamefont {Illing}\ and\ \citenamefont
  {Gauthier}(2005)}]{IllingPD05}%
  \BibitemOpen
  \bibfield  {author} {\bibinfo {author} {\bibfnamefont {L.}~\bibnamefont
  {Illing}}\ and\ \bibinfo {author} {\bibfnamefont {D.~J.}\ \bibnamefont
  {Gauthier}},\ }\href {\doibase 10.1016/j.physd.2005.07.008} {\bibfield
  {journal} {\bibinfo  {journal} {Physica D}\ }\textbf {\bibinfo {volume}
  {210}},\ \bibinfo {pages} {180} (\bibinfo {year} {2005})}\BibitemShut
  {NoStop}%
\bibitem [{\citenamefont {Kouomou}\ \emph {et~al.}(2005)\citenamefont
  {Kouomou}, \citenamefont {Colet}, \citenamefont {Larger},\ and\ \citenamefont
  {Gastaud}}]{KouomouPRL05}%
  \BibitemOpen
  \bibfield  {author} {\bibinfo {author} {\bibfnamefont {Y.~C.}\ \bibnamefont
  {Kouomou}}, \bibinfo {author} {\bibfnamefont {P.}~\bibnamefont {Colet}},
  \bibinfo {author} {\bibfnamefont {L.}~\bibnamefont {Larger}}, \ and\ \bibinfo
  {author} {\bibfnamefont {N.}~\bibnamefont {Gastaud}},\ }\href {\doibase
  10.1103/physrevlett.95.203903} {\bibfield  {journal} {\bibinfo  {journal}
  {Phys. Rev. Lett.}\ }\textbf {\bibinfo {volume} {95}} (\bibinfo {year}
  {2005}),\ 10.1103/physrevlett.95.203903}\BibitemShut {NoStop}%
\bibitem [{\citenamefont {Soriano}\ \emph {et~al.}(2013)\citenamefont
  {Soriano}, \citenamefont {Garc{\'{\i}}a-Ojalvo}, \citenamefont {Mirasso},\
  and\ \citenamefont {Fischer}}]{SorianoRMP13}%
  \BibitemOpen
  \bibfield  {author} {\bibinfo {author} {\bibfnamefont {M.~C.}\ \bibnamefont
  {Soriano}}, \bibinfo {author} {\bibfnamefont {J.}~\bibnamefont
  {Garc{\'{\i}}a-Ojalvo}}, \bibinfo {author} {\bibfnamefont {C.~R.}\
  \bibnamefont {Mirasso}}, \ and\ \bibinfo {author} {\bibfnamefont
  {I.}~\bibnamefont {Fischer}},\ }\href {\doibase 10.1103/revmodphys.85.421}
  {\bibfield  {journal} {\bibinfo  {journal} {Rev. Mod. Phys.}\ }\textbf
  {\bibinfo {volume} {85}},\ \bibinfo {pages} {421} (\bibinfo {year}
  {2013})}\BibitemShut {NoStop}%
\bibitem [{\citenamefont {Hutchinson}(1948)}]{HutchinsonANAS48}%
  \BibitemOpen
  \bibfield  {author} {\bibinfo {author} {\bibfnamefont {G.~E.}\ \bibnamefont
  {Hutchinson}},\ }\href {\doibase 10.1111/j.1749-6632.1948.tb39854.x}
  {\bibfield  {journal} {\bibinfo  {journal} {Ann. N.Y. Acad. Sci.}\ }\textbf
  {\bibinfo {volume} {50}},\ \bibinfo {pages} {221} (\bibinfo {year}
  {1948})}\BibitemShut {NoStop}%
\bibitem [{\citenamefont {Epstein}\ and\ \citenamefont
  {Luo}(1991)}]{EpsteinJCP91}%
  \BibitemOpen
  \bibfield  {author} {\bibinfo {author} {\bibfnamefont {I.~R.}\ \bibnamefont
  {Epstein}}\ and\ \bibinfo {author} {\bibfnamefont {Y.}~\bibnamefont {Luo}},\
  }\href {\doibase 10.1063/1.461481} {\bibfield  {journal} {\bibinfo  {journal}
  {J. Chem. Phys.}\ }\textbf {\bibinfo {volume} {95}},\ \bibinfo {pages} {244}
  (\bibinfo {year} {1991})}\BibitemShut {NoStop}%
\bibitem [{\citenamefont {Roussel}(1996)}]{RousselJPC96}%
  \BibitemOpen
  \bibfield  {author} {\bibinfo {author} {\bibfnamefont {M.~R.}\ \bibnamefont
  {Roussel}},\ }\href {\doibase 10.1021/jp9600672} {\bibfield  {journal}
  {\bibinfo  {journal} {J. Phys. Chem.}\ }\textbf {\bibinfo {volume} {100}},\
  \bibinfo {pages} {8323} (\bibinfo {year} {1996})}\BibitemShut {NoStop}%
\bibitem [{\citenamefont {Foss}\ \emph {et~al.}(1996)\citenamefont {Foss},
  \citenamefont {Longtin}, \citenamefont {Mensour},\ and\ \citenamefont
  {Milton}}]{FossPRL96}%
  \BibitemOpen
  \bibfield  {author} {\bibinfo {author} {\bibfnamefont {J.}~\bibnamefont
  {Foss}}, \bibinfo {author} {\bibfnamefont {A.}~\bibnamefont {Longtin}},
  \bibinfo {author} {\bibfnamefont {B.}~\bibnamefont {Mensour}}, \ and\
  \bibinfo {author} {\bibfnamefont {J.}~\bibnamefont {Milton}},\ }\href
  {\doibase 10.1103/physrevlett.76.708} {\bibfield  {journal} {\bibinfo
  {journal} {Phys. Rev. Lett.}\ }\textbf {\bibinfo {volume} {76}},\ \bibinfo
  {pages} {708} (\bibinfo {year} {1996})}\BibitemShut {NoStop}%
\bibitem [{\citenamefont {Hizanidis}\ \emph {et~al.}(2008)\citenamefont
  {Hizanidis}, \citenamefont {Aust},\ and\ \citenamefont
  {Sch\"oll}}]{HizanidisIJBC08}%
  \BibitemOpen
  \bibfield  {author} {\bibinfo {author} {\bibfnamefont {J.}~\bibnamefont
  {Hizanidis}}, \bibinfo {author} {\bibfnamefont {R.}~\bibnamefont {Aust}}, \
  and\ \bibinfo {author} {\bibfnamefont {E.}~\bibnamefont {Sch\"oll}},\ }\href
  {\doibase 10.1142/s0218127408021348} {\bibfield  {journal} {\bibinfo
  {journal} {Int. J. Bifurcation Chaos}\ }\textbf {\bibinfo {volume} {18}},\
  \bibinfo {pages} {1759} (\bibinfo {year} {2008})}\BibitemShut {NoStop}%
\bibitem [{\citenamefont {Prucnal}\ \emph {et~al.}(2016)\citenamefont
  {Prucnal}, \citenamefont {Shastri}, \citenamefont {de~Lima}, \citenamefont
  {Nahmias},\ and\ \citenamefont {Tait}}]{prucnal2016recent}%
  \BibitemOpen
  \bibfield  {author} {\bibinfo {author} {\bibfnamefont {P.~R.}\ \bibnamefont
  {Prucnal}}, \bibinfo {author} {\bibfnamefont {B.~J.}\ \bibnamefont
  {Shastri}}, \bibinfo {author} {\bibfnamefont {T.~F.}\ \bibnamefont
  {de~Lima}}, \bibinfo {author} {\bibfnamefont {M.~A.}\ \bibnamefont
  {Nahmias}}, \ and\ \bibinfo {author} {\bibfnamefont {A.~N.}\ \bibnamefont
  {Tait}},\ }\href@noop {} {\bibfield  {journal} {\bibinfo  {journal} {Advances
  in Optics and Photonics}\ }\textbf {\bibinfo {volume} {8}},\ \bibinfo {pages}
  {228} (\bibinfo {year} {2016})}\BibitemShut {NoStop}%
\bibitem [{\citenamefont {Nahmias}\ \emph {et~al.}(2013)\citenamefont
  {Nahmias}, \citenamefont {Shastri}, \citenamefont {Tait},\ and\ \citenamefont
  {Prucnal}}]{nahmias2013leaky}%
  \BibitemOpen
  \bibfield  {author} {\bibinfo {author} {\bibfnamefont {M.~A.}\ \bibnamefont
  {Nahmias}}, \bibinfo {author} {\bibfnamefont {B.~J.}\ \bibnamefont
  {Shastri}}, \bibinfo {author} {\bibfnamefont {A.~N.}\ \bibnamefont {Tait}}, \
  and\ \bibinfo {author} {\bibfnamefont {P.~R.}\ \bibnamefont {Prucnal}},\
  }\href@noop {} {\bibfield  {journal} {\bibinfo  {journal} {IEEE journal of
  selected topics in quantum electronics}\ }\textbf {\bibinfo {volume} {19}},\
  \bibinfo {pages} {1} (\bibinfo {year} {2013})}\BibitemShut {NoStop}%
\bibitem [{\citenamefont {Terrien}\ \emph {et~al.}(2020)\citenamefont
  {Terrien}, \citenamefont {Pammi}, \citenamefont {Broderick}, \citenamefont
  {Braive}, \citenamefont {Beaudoin}, \citenamefont {Sagnes}, \citenamefont
  {Krauskopf},\ and\ \citenamefont {Barbay}}]{TerrienPRR20}%
  \BibitemOpen
  \bibfield  {author} {\bibinfo {author} {\bibfnamefont {S.}~\bibnamefont
  {Terrien}}, \bibinfo {author} {\bibfnamefont {V.~A.}\ \bibnamefont {Pammi}},
  \bibinfo {author} {\bibfnamefont {N.~G.~R.}\ \bibnamefont {Broderick}},
  \bibinfo {author} {\bibfnamefont {R.}~\bibnamefont {Braive}}, \bibinfo
  {author} {\bibfnamefont {G.}~\bibnamefont {Beaudoin}}, \bibinfo {author}
  {\bibfnamefont {I.}~\bibnamefont {Sagnes}}, \bibinfo {author} {\bibfnamefont
  {B.}~\bibnamefont {Krauskopf}}, \ and\ \bibinfo {author} {\bibfnamefont
  {S.}~\bibnamefont {Barbay}},\ }\href {\doibase
  10.1103/physrevresearch.2.023012} {\bibfield  {journal} {\bibinfo  {journal}
  {Phys. Rev. Research}\ }\textbf {\bibinfo {volume} {2}} (\bibinfo {year}
  {2020}),\ 10.1103/physrevresearch.2.023012}\BibitemShut {NoStop}%
\bibitem [{\citenamefont {Garbin}\ \emph {et~al.}(2015)\citenamefont {Garbin},
  \citenamefont {Javaloyes}, \citenamefont {Tissoni},\ and\ \citenamefont
  {Barland}}]{GarbinNC15}%
  \BibitemOpen
  \bibfield  {author} {\bibinfo {author} {\bibfnamefont {B.}~\bibnamefont
  {Garbin}}, \bibinfo {author} {\bibfnamefont {J.}~\bibnamefont {Javaloyes}},
  \bibinfo {author} {\bibfnamefont {G.}~\bibnamefont {Tissoni}}, \ and\
  \bibinfo {author} {\bibfnamefont {S.}~\bibnamefont {Barland}},\ }\href
  {http://dx.doi.org/10.1038/ncomms6915} {\bibfield  {journal} {\bibinfo
  {journal} {Nat. Commun.}\ }\textbf {\bibinfo {volume} {6}},\  (\bibinfo
  {year} {2015})}\BibitemShut {NoStop}%
\bibitem [{\citenamefont {Romeira}\ \emph {et~al.}(2016)\citenamefont
  {Romeira}, \citenamefont {Av\'o}, \citenamefont {Figueiredo}, \citenamefont
  {Barland},\ and\ \citenamefont {Javaloyes}}]{RomeiraNSR16}%
  \BibitemOpen
  \bibfield  {author} {\bibinfo {author} {\bibfnamefont {B.}~\bibnamefont
  {Romeira}}, \bibinfo {author} {\bibfnamefont {R.}~\bibnamefont {Av\'o}},
  \bibinfo {author} {\bibfnamefont {J.~M.~L.}\ \bibnamefont {Figueiredo}},
  \bibinfo {author} {\bibfnamefont {S.}~\bibnamefont {Barland}}, \ and\
  \bibinfo {author} {\bibfnamefont {J.}~\bibnamefont {Javaloyes}},\ }\href@noop
  {} {\bibfield  {journal} {\bibinfo  {journal} {Sci. Rep.}\ }\textbf {\bibinfo
  {volume} {6}} (\bibinfo {year} {2016})}\BibitemShut {NoStop}%
\bibitem [{\citenamefont {Terrien}\ \emph
  {et~al.}(2017{\natexlab{a}})\citenamefont {Terrien}, \citenamefont
  {Krauskopf}, \citenamefont {Broderick}, \citenamefont {Andr{\'{e}}oli},
  \citenamefont {Selmi}, \citenamefont {Braive}, \citenamefont {Beaudoin},
  \citenamefont {Sagnes},\ and\ \citenamefont {Barbay}}]{TerrienPRA17}%
  \BibitemOpen
  \bibfield  {author} {\bibinfo {author} {\bibfnamefont {S.}~\bibnamefont
  {Terrien}}, \bibinfo {author} {\bibfnamefont {B.}~\bibnamefont {Krauskopf}},
  \bibinfo {author} {\bibfnamefont {N.~G.~R.}\ \bibnamefont {Broderick}},
  \bibinfo {author} {\bibfnamefont {L.}~\bibnamefont {Andr{\'{e}}oli}},
  \bibinfo {author} {\bibfnamefont {F.}~\bibnamefont {Selmi}}, \bibinfo
  {author} {\bibfnamefont {R.}~\bibnamefont {Braive}}, \bibinfo {author}
  {\bibfnamefont {G.}~\bibnamefont {Beaudoin}}, \bibinfo {author}
  {\bibfnamefont {I.}~\bibnamefont {Sagnes}}, \ and\ \bibinfo {author}
  {\bibfnamefont {S.}~\bibnamefont {Barbay}},\ }\href {\doibase
  10.1103/physreva.96.043863} {\bibfield  {journal} {\bibinfo  {journal} {Phys.
  Rev. A}\ }\textbf {\bibinfo {volume} {96}} (\bibinfo {year}
  {2017}{\natexlab{a}}),\ 10.1103/physreva.96.043863}\BibitemShut {NoStop}%
\bibitem [{\citenamefont {Terrien}\ \emph {et~al.}(2018)\citenamefont
  {Terrien}, \citenamefont {Krauskopf}, \citenamefont {Broderick},
  \citenamefont {Braive}, \citenamefont {Beaudoin}, \citenamefont {Sagnes},\
  and\ \citenamefont {Barbay}}]{TerrienOL18}%
  \BibitemOpen
  \bibfield  {author} {\bibinfo {author} {\bibfnamefont {S.}~\bibnamefont
  {Terrien}}, \bibinfo {author} {\bibfnamefont {B.}~\bibnamefont {Krauskopf}},
  \bibinfo {author} {\bibfnamefont {N.~G.}\ \bibnamefont {Broderick}}, \bibinfo
  {author} {\bibfnamefont {R.}~\bibnamefont {Braive}}, \bibinfo {author}
  {\bibfnamefont {G.}~\bibnamefont {Beaudoin}}, \bibinfo {author}
  {\bibfnamefont {I.}~\bibnamefont {Sagnes}}, \ and\ \bibinfo {author}
  {\bibfnamefont {S.}~\bibnamefont {Barbay}},\ }\href@noop {} {\bibfield
  {journal} {\bibinfo  {journal} {Opt. Lett.}\ }\textbf {\bibinfo {volume}
  {43}},\ \bibinfo {pages} {3013} (\bibinfo {year} {2018})}\BibitemShut
  {NoStop}%
\bibitem [{\citenamefont {Selmi}\ \emph {et~al.}(2016)\citenamefont {Selmi},
  \citenamefont {Braive}, \citenamefont {Beaudoin}, \citenamefont {Sagnes},
  \citenamefont {Kuszelewicz}, \citenamefont {Erneux},\ and\ \citenamefont
  {Barbay}}]{SelmiPRE16}%
  \BibitemOpen
  \bibfield  {author} {\bibinfo {author} {\bibfnamefont {F.}~\bibnamefont
  {Selmi}}, \bibinfo {author} {\bibfnamefont {R.}~\bibnamefont {Braive}},
  \bibinfo {author} {\bibfnamefont {G.}~\bibnamefont {Beaudoin}}, \bibinfo
  {author} {\bibfnamefont {I.}~\bibnamefont {Sagnes}}, \bibinfo {author}
  {\bibfnamefont {R.}~\bibnamefont {Kuszelewicz}}, \bibinfo {author}
  {\bibfnamefont {T.}~\bibnamefont {Erneux}}, \ and\ \bibinfo {author}
  {\bibfnamefont {S.}~\bibnamefont {Barbay}},\ }\href {\doibase
  10.1103/PhysRevE.94.042219} {\bibfield  {journal} {\bibinfo  {journal} {Phys.
  Rev. E}\ }\textbf {\bibinfo {volume} {94}},\ \bibinfo {pages} {042219}
  (\bibinfo {year} {2016})}\BibitemShut {NoStop}%
\bibitem [{\citenamefont {Erneux}\ and\ \citenamefont
  {Barbay}(2018)}]{ErneuxPRE18}%
  \BibitemOpen
  \bibfield  {author} {\bibinfo {author} {\bibfnamefont {T.}~\bibnamefont
  {Erneux}}\ and\ \bibinfo {author} {\bibfnamefont {S.}~\bibnamefont
  {Barbay}},\ }\href {\doibase 10.1103/physreve.97.062214} {\bibfield
  {journal} {\bibinfo  {journal} {Phys. Rev. E}\ }\textbf {\bibinfo {volume}
  {97}} (\bibinfo {year} {2018}),\ 10.1103/physreve.97.062214}\BibitemShut
  {NoStop}%
\bibitem [{\citenamefont {Yamada}(1993)}]{Yamada93}%
  \BibitemOpen
  \bibfield  {author} {\bibinfo {author} {\bibfnamefont {M.}~\bibnamefont
  {Yamada}},\ }\href@noop {} {\bibfield  {journal} {\bibinfo  {journal} {IEEE
  J. Quantum Electron.}\ }\textbf {\bibinfo {volume} {29}},\ \bibinfo {pages}
  {1330} (\bibinfo {year} {1993})}\BibitemShut {NoStop}%
\bibitem [{\citenamefont {Krauskopf}\ and\ \citenamefont
  {Walker}(2012)}]{KrauskopfWalker}%
  \BibitemOpen
  \bibfield  {author} {\bibinfo {author} {\bibfnamefont {B.}~\bibnamefont
  {Krauskopf}}\ and\ \bibinfo {author} {\bibfnamefont {J.~J.}\ \bibnamefont
  {Walker}},\ }\enquote {\bibinfo {title} {Bifurcation study of a semiconductor
  laser with saturable absorber and delayed optical feedback},}\ in\ \href@noop
  {} {\emph {\bibinfo {booktitle} {Nonlinear Laser Dynamics}}}\ (\bibinfo
  {publisher} {Wiley-VCH Verlag GmbH \& Co. KGaA},\ \bibinfo {year} {2012})\
  pp.\ \bibinfo {pages} {161--181}\BibitemShut {NoStop}%
\bibitem [{\citenamefont {Barbay}\ \emph {et~al.}(2011)\citenamefont {Barbay},
  \citenamefont {Kuszelewicz},\ and\ \citenamefont {Yacomotti}}]{BarbayOL11}%
  \BibitemOpen
  \bibfield  {author} {\bibinfo {author} {\bibfnamefont {S.}~\bibnamefont
  {Barbay}}, \bibinfo {author} {\bibfnamefont {R.}~\bibnamefont {Kuszelewicz}},
  \ and\ \bibinfo {author} {\bibfnamefont {A.~M.}\ \bibnamefont {Yacomotti}},\
  }\href {\doibase 10.1364/OL.36.004476} {\bibfield  {journal} {\bibinfo
  {journal} {Opt. Lett.}\ }\textbf {\bibinfo {volume} {36}},\ \bibinfo {pages}
  {4476} (\bibinfo {year} {2011})}\BibitemShut {NoStop}%
\bibitem [{\citenamefont {Terrien}\ \emph
  {et~al.}(2017{\natexlab{b}})\citenamefont {Terrien}, \citenamefont
  {Krauskopf},\ and\ \citenamefont {Broderick}}]{TerrienSIADS17}%
  \BibitemOpen
  \bibfield  {author} {\bibinfo {author} {\bibfnamefont {S.}~\bibnamefont
  {Terrien}}, \bibinfo {author} {\bibfnamefont {B.}~\bibnamefont {Krauskopf}},
  \ and\ \bibinfo {author} {\bibfnamefont {N.~G.~R.}\ \bibnamefont
  {Broderick}},\ }\href@noop {} {\bibfield  {journal} {\bibinfo  {journal}
  {SIAM J. Appl. Dyn. Sys.}\ }\textbf {\bibinfo {volume} {16}},\ \bibinfo
  {pages} {771} (\bibinfo {year} {2017}{\natexlab{b}})}\BibitemShut {NoStop}%
\bibitem [{\citenamefont {Kuznetsov}(2013)}]{Kuznetsov2013elements}%
  \BibitemOpen
  \bibfield  {author} {\bibinfo {author} {\bibfnamefont {Y.~A.}\ \bibnamefont
  {Kuznetsov}},\ }\href@noop {} {\emph {\bibinfo {title} {Elements of applied
  bifurcation theory}}},\ Vol.\ \bibinfo {volume} {112}\ (\bibinfo  {publisher}
  {Springer Science \& Business Media},\ \bibinfo {year} {2013})\BibitemShut
  {NoStop}%
\bibitem [{\citenamefont {Yanchuk}\ \emph {et~al.}(2019)\citenamefont
  {Yanchuk}, \citenamefont {Ruschel}, \citenamefont {Sieber},\ and\
  \citenamefont {Wolfrum}}]{YanchukPRL19}%
  \BibitemOpen
  \bibfield  {author} {\bibinfo {author} {\bibfnamefont {S.}~\bibnamefont
  {Yanchuk}}, \bibinfo {author} {\bibfnamefont {S.}~\bibnamefont {Ruschel}},
  \bibinfo {author} {\bibfnamefont {J.}~\bibnamefont {Sieber}}, \ and\ \bibinfo
  {author} {\bibfnamefont {M.}~\bibnamefont {Wolfrum}},\ }\href {\doibase
  10.1103/physrevlett.123.053901} {\bibfield  {journal} {\bibinfo  {journal}
  {Phys. Rev. Lett.}\ }\textbf {\bibinfo {volume} {123}} (\bibinfo {year}
  {2019}),\ 10.1103/physrevlett.123.053901}\BibitemShut {NoStop}%
\bibitem [{\citenamefont {Elsass}\ \emph {et~al.}(2010)\citenamefont {Elsass},
  \citenamefont {Gauthron}, \citenamefont {Beaudoin}, \citenamefont {Sagnes},
  \citenamefont {Kuszelewicz},\ and\ \citenamefont {Barbay}}]{ElsassEPJD10}%
  \BibitemOpen
  \bibfield  {author} {\bibinfo {author} {\bibfnamefont {T.}~\bibnamefont
  {Elsass}}, \bibinfo {author} {\bibfnamefont {K.}~\bibnamefont {Gauthron}},
  \bibinfo {author} {\bibfnamefont {G.}~\bibnamefont {Beaudoin}}, \bibinfo
  {author} {\bibfnamefont {I.}~\bibnamefont {Sagnes}}, \bibinfo {author}
  {\bibfnamefont {R.}~\bibnamefont {Kuszelewicz}}, \ and\ \bibinfo {author}
  {\bibfnamefont {S.}~\bibnamefont {Barbay}},\ }\href
  {http://dx.doi.org/10.1140/epjd/e2010-00079-6} {\bibfield  {journal}
  {\bibinfo  {journal} {Eur. Phys. J. D}\ }\textbf {\bibinfo {volume} {59}}
  (\bibinfo {year} {2010})}\BibitemShut {NoStop}%
\bibitem [{\citenamefont {Selmi}\ \emph {et~al.}(2014)\citenamefont {Selmi},
  \citenamefont {Braive}, \citenamefont {Beaudoin}, \citenamefont {Sagnes},
  \citenamefont {Kuszelewicz},\ and\ \citenamefont {Barbay}}]{SelmiPRL14}%
  \BibitemOpen
  \bibfield  {author} {\bibinfo {author} {\bibfnamefont {F.}~\bibnamefont
  {Selmi}}, \bibinfo {author} {\bibfnamefont {R.}~\bibnamefont {Braive}},
  \bibinfo {author} {\bibfnamefont {G.}~\bibnamefont {Beaudoin}}, \bibinfo
  {author} {\bibfnamefont {I.}~\bibnamefont {Sagnes}}, \bibinfo {author}
  {\bibfnamefont {R.}~\bibnamefont {Kuszelewicz}}, \ and\ \bibinfo {author}
  {\bibfnamefont {S.}~\bibnamefont {Barbay}},\ }\href {\doibase
  10.1103/PhysRevLett.112.183902} {\bibfield  {journal} {\bibinfo  {journal}
  {Phys. Rev. Lett.}\ }\textbf {\bibinfo {volume} {112}},\ \bibinfo {pages}
  {183902} (\bibinfo {year} {2014})}\BibitemShut {NoStop}%
\bibitem [{\citenamefont {Dubbeldam}\ \emph {et~al.}(1999)\citenamefont
  {Dubbeldam}, \citenamefont {Krauskopf},\ and\ \citenamefont
  {Lenstra}}]{DubbeldamPRE99}%
  \BibitemOpen
  \bibfield  {author} {\bibinfo {author} {\bibfnamefont {J.~L.~A.}\
  \bibnamefont {Dubbeldam}}, \bibinfo {author} {\bibfnamefont {B.}~\bibnamefont
  {Krauskopf}}, \ and\ \bibinfo {author} {\bibfnamefont {D.}~\bibnamefont
  {Lenstra}},\ }\href {\doibase 10.1103/PhysRevE.60.6580} {\bibfield  {journal}
  {\bibinfo  {journal} {Phys. Rev. E}\ }\textbf {\bibinfo {volume} {60}},\
  \bibinfo {pages} {6580} (\bibinfo {year} {1999})}\BibitemShut {NoStop}%
\bibitem [{\citenamefont {Izhikevich}(2007)}]{IzhikevichBook}%
  \BibitemOpen
  \bibfield  {author} {\bibinfo {author} {\bibfnamefont {E.}~\bibnamefont
  {Izhikevich}},\ }\href@noop {} {\emph {\bibinfo {title} {Dynamical Systems in
  Neuroscience: The Geometry of Excitability and Bursting.}}}\ (\bibinfo
  {publisher} {The MIT press},\ \bibinfo {year} {2007})\BibitemShut {NoStop}%
\bibitem [{\citenamefont {Dubbeldam}\ and\ \citenamefont
  {Krauskopf}(1999)}]{DubbeldamOC99}%
  \BibitemOpen
  \bibfield  {author} {\bibinfo {author} {\bibfnamefont {J.~L.~A.}\
  \bibnamefont {Dubbeldam}}\ and\ \bibinfo {author} {\bibfnamefont
  {B.}~\bibnamefont {Krauskopf}},\ }\href@noop {} {\bibfield  {journal}
  {\bibinfo  {journal} {Opt. Commun.}\ }\textbf {\bibinfo {volume} {159}},\
  \bibinfo {pages} {325} (\bibinfo {year} {1999})}\BibitemShut {NoStop}%
\bibitem [{\citenamefont {Otupiri}\ \emph {et~al.}(2019)\citenamefont
  {Otupiri}, \citenamefont {Krauskopf},\ and\ \citenamefont
  {Broderick}}]{otupiri2019yamada}%
  \BibitemOpen
  \bibfield  {author} {\bibinfo {author} {\bibfnamefont {R.}~\bibnamefont
  {Otupiri}}, \bibinfo {author} {\bibfnamefont {B.}~\bibnamefont {Krauskopf}},
  \ and\ \bibinfo {author} {\bibfnamefont {N.~G.}\ \bibnamefont {Broderick}},\
  }\href@noop {} {\bibfield  {journal} {\bibinfo  {journal} {arXiv preprint
  arXiv:1911.01835}\ } (\bibinfo {year} {2019})}\BibitemShut {NoStop}%
\bibitem [{\citenamefont {Vreeswijk}\ \emph {et~al.}(1994)\citenamefont
  {Vreeswijk}, \citenamefont {Abbott},\ and\ \citenamefont
  {Ermentrout}}]{VreeswijkJCN94}%
  \BibitemOpen
  \bibfield  {author} {\bibinfo {author} {\bibfnamefont {C.~V.}\ \bibnamefont
  {Vreeswijk}}, \bibinfo {author} {\bibfnamefont {L.~F.}\ \bibnamefont
  {Abbott}}, \ and\ \bibinfo {author} {\bibfnamefont {G.~B.}\ \bibnamefont
  {Ermentrout}},\ }\href {\doibase 10.1007/bf00961879} {\bibfield  {journal}
  {\bibinfo  {journal} {J. Comput. Neurosci.}\ }\textbf {\bibinfo {volume}
  {1}},\ \bibinfo {pages} {313} (\bibinfo {year} {1994})}\BibitemShut {NoStop}%
\bibitem [{\citenamefont {Bressloff}\ and\ \citenamefont
  {Coombes}(2000)}]{BressloffSJAM00}%
  \BibitemOpen
  \bibfield  {author} {\bibinfo {author} {\bibfnamefont {P.~C.}\ \bibnamefont
  {Bressloff}}\ and\ \bibinfo {author} {\bibfnamefont {S.}~\bibnamefont
  {Coombes}},\ }\href {\doibase 10.1137/s0036139998339643} {\bibfield
  {journal} {\bibinfo  {journal} {{SIAM} J. Appl. Math.}\ }\textbf {\bibinfo
  {volume} {60}},\ \bibinfo {pages} {820} (\bibinfo {year} {2000})}\BibitemShut
  {NoStop}%
\bibitem [{\citenamefont {Klinshov}\ \emph
  {et~al.}(2015{\natexlab{a}})\citenamefont {Klinshov}, \citenamefont
  {L{\"u}cken}, \citenamefont {Shchapin}, \citenamefont {Nekorkin},\ and\
  \citenamefont {Yanchuk}}]{KlinshovPRL15}%
  \BibitemOpen
  \bibfield  {author} {\bibinfo {author} {\bibfnamefont {V.}~\bibnamefont
  {Klinshov}}, \bibinfo {author} {\bibfnamefont {L.}~\bibnamefont
  {L{\"u}cken}}, \bibinfo {author} {\bibfnamefont {D.}~\bibnamefont
  {Shchapin}}, \bibinfo {author} {\bibfnamefont {V.}~\bibnamefont {Nekorkin}},
  \ and\ \bibinfo {author} {\bibfnamefont {S.}~\bibnamefont {Yanchuk}},\ }\href
  {\doibase 10.1103/physrevlett.114.178103} {\bibfield  {journal} {\bibinfo
  {journal} {Phys. Rev. Lett.}\ }\textbf {\bibinfo {volume} {114}} (\bibinfo
  {year} {2015}{\natexlab{a}}),\ 10.1103/physrevlett.114.178103}\BibitemShut
  {NoStop}%
\bibitem [{\citenamefont {Klinshov}\ \emph
  {et~al.}(2015{\natexlab{b}})\citenamefont {Klinshov}, \citenamefont
  {L{\"u}cken}, \citenamefont {Shchapin}, \citenamefont {Nekorkin},\ and\
  \citenamefont {Yanchuk}}]{KlinshovPRE15}%
  \BibitemOpen
  \bibfield  {author} {\bibinfo {author} {\bibfnamefont {V.}~\bibnamefont
  {Klinshov}}, \bibinfo {author} {\bibfnamefont {L.}~\bibnamefont
  {L{\"u}cken}}, \bibinfo {author} {\bibfnamefont {D.}~\bibnamefont
  {Shchapin}}, \bibinfo {author} {\bibfnamefont {V.}~\bibnamefont {Nekorkin}},
  \ and\ \bibinfo {author} {\bibfnamefont {S.}~\bibnamefont {Yanchuk}},\ }\href
  {\doibase 10.1103/physreve.92.042914} {\bibfield  {journal} {\bibinfo
  {journal} {Physical Review E}\ }\textbf {\bibinfo {volume} {92}} (\bibinfo
  {year} {2015}{\natexlab{b}}),\ 10.1103/physreve.92.042914}\BibitemShut
  {NoStop}%
\bibitem [{\citenamefont {Grelu}\ and\ \citenamefont
  {Soto-Crespo}(2008)}]{Grelu08}%
  \BibitemOpen
  \bibfield  {author} {\bibinfo {author} {\bibfnamefont {P.}~\bibnamefont
  {Grelu}}\ and\ \bibinfo {author} {\bibfnamefont {J.}~\bibnamefont
  {Soto-Crespo}},\ }in\ \href {\doibase 10.1007/978-3-540-78217-9_6} {\emph
  {\bibinfo {booktitle} {Lecture Notes in Physics}}}\ (\bibinfo  {publisher}
  {Springer Berlin Heidelberg},\ \bibinfo {year} {2008})\ pp.\ \bibinfo {pages}
  {1--37}\BibitemShut {NoStop}%
\bibitem [{\citenamefont {Krupa}\ \emph {et~al.}(2017)\citenamefont {Krupa},
  \citenamefont {Nithyanandan}, \citenamefont {Andral}, \citenamefont
  {Tchofo-Dinda},\ and\ \citenamefont {Grelu}}]{KrupaPRL17}%
  \BibitemOpen
  \bibfield  {author} {\bibinfo {author} {\bibfnamefont {K.}~\bibnamefont
  {Krupa}}, \bibinfo {author} {\bibfnamefont {K.}~\bibnamefont {Nithyanandan}},
  \bibinfo {author} {\bibfnamefont {U.}~\bibnamefont {Andral}}, \bibinfo
  {author} {\bibfnamefont {P.}~\bibnamefont {Tchofo-Dinda}}, \ and\ \bibinfo
  {author} {\bibfnamefont {P.}~\bibnamefont {Grelu}},\ }\href {\doibase
  10.1103/physrevlett.118.243901} {\bibfield  {journal} {\bibinfo  {journal}
  {Phys. Rev. Lett.}\ }\textbf {\bibinfo {volume} {118}} (\bibinfo {year}
  {2017}),\ 10.1103/physrevlett.118.243901}\BibitemShut {NoStop}%
\bibitem [{\citenamefont {Aadhi}\ \emph {et~al.}(2019)\citenamefont {Aadhi},
  \citenamefont {Kovalev}, \citenamefont {Kues}, \citenamefont {Roztocki},
  \citenamefont {Reimer}, \citenamefont {Zhang}, \citenamefont {Wang},
  \citenamefont {Little}, \citenamefont {Chu}, \citenamefont {Wang},
  \citenamefont {Moss}, \citenamefont {Viktorov},\ and\ \citenamefont
  {Morandotti}}]{AadhiOE19}%
  \BibitemOpen
  \bibfield  {author} {\bibinfo {author} {\bibfnamefont {A.}~\bibnamefont
  {Aadhi}}, \bibinfo {author} {\bibfnamefont {A.~V.}\ \bibnamefont {Kovalev}},
  \bibinfo {author} {\bibfnamefont {M.}~\bibnamefont {Kues}}, \bibinfo {author}
  {\bibfnamefont {P.}~\bibnamefont {Roztocki}}, \bibinfo {author}
  {\bibfnamefont {C.}~\bibnamefont {Reimer}}, \bibinfo {author} {\bibfnamefont
  {Y.}~\bibnamefont {Zhang}}, \bibinfo {author} {\bibfnamefont
  {T.}~\bibnamefont {Wang}}, \bibinfo {author} {\bibfnamefont {B.~E.}\
  \bibnamefont {Little}}, \bibinfo {author} {\bibfnamefont {S.~T.}\
  \bibnamefont {Chu}}, \bibinfo {author} {\bibfnamefont {Z.}~\bibnamefont
  {Wang}}, \bibinfo {author} {\bibfnamefont {D.~J.}\ \bibnamefont {Moss}},
  \bibinfo {author} {\bibfnamefont {E.~A.}\ \bibnamefont {Viktorov}}, \ and\
  \bibinfo {author} {\bibfnamefont {R.}~\bibnamefont {Morandotti}},\ }\href
  {\doibase 10.1364/oe.27.025251} {\bibfield  {journal} {\bibinfo  {journal}
  {Optics Express}\ }\textbf {\bibinfo {volume} {27}},\ \bibinfo {pages}
  {25251} (\bibinfo {year} {2019})}\BibitemShut {NoStop}%
\bibitem [{\citenamefont {Romeira}\ \emph {et~al.}(2017)\citenamefont
  {Romeira}, \citenamefont {Figueiredo},\ and\ \citenamefont
  {Javaloyes}}]{RomeiraC17}%
  \BibitemOpen
  \bibfield  {author} {\bibinfo {author} {\bibfnamefont {B.}~\bibnamefont
  {Romeira}}, \bibinfo {author} {\bibfnamefont {J.~M.~L.}\ \bibnamefont
  {Figueiredo}}, \ and\ \bibinfo {author} {\bibfnamefont {J.}~\bibnamefont
  {Javaloyes}},\ }\href {\doibase 10.1063/1.5008888} {\bibfield  {journal}
  {\bibinfo  {journal} {Chaos}\ }\textbf {\bibinfo {volume} {27}},\ \bibinfo
  {pages} {114323} (\bibinfo {year} {2017})}\BibitemShut {NoStop}%
\bibitem [{\citenamefont {{Peng}}\ \emph {et~al.}(2019)\citenamefont {{Peng}},
  \citenamefont {{de Lima}}, \citenamefont {{Nahmias}}, \citenamefont {{Tait}},
  \citenamefont {{Shastri}},\ and\ \citenamefont {{Prucnal}}}]{Peng19}%
  \BibitemOpen
  \bibfield  {author} {\bibinfo {author} {\bibfnamefont {H.}~\bibnamefont
  {{Peng}}}, \bibinfo {author} {\bibfnamefont {T.~F.}\ \bibnamefont {{de
  Lima}}}, \bibinfo {author} {\bibfnamefont {M.~A.}\ \bibnamefont {{Nahmias}}},
  \bibinfo {author} {\bibfnamefont {A.~N.}\ \bibnamefont {{Tait}}}, \bibinfo
  {author} {\bibfnamefont {B.~J.}\ \bibnamefont {{Shastri}}}, \ and\ \bibinfo
  {author} {\bibfnamefont {P.~R.}\ \bibnamefont {{Prucnal}}},\ }in\ \href@noop
  {} {\emph {\bibinfo {booktitle} {2019 Conference on Lasers and Electro-Optics
  (CLEO)}}}\ (\bibinfo {year} {2019})\ pp.\ \bibinfo {pages} {1--2}\BibitemShut
  {NoStop}%
\bibitem [{\citenamefont {Pammi}\ \emph {et~al.}(2020)\citenamefont {Pammi},
  \citenamefont {Alfaro-Bittner}, \citenamefont {Clerc},\ and\ \citenamefont
  {Barbay}}]{PammiIJSTQE20}%
  \BibitemOpen
  \bibfield  {author} {\bibinfo {author} {\bibfnamefont {V.~A.}\ \bibnamefont
  {Pammi}}, \bibinfo {author} {\bibfnamefont {K.}~\bibnamefont
  {Alfaro-Bittner}}, \bibinfo {author} {\bibfnamefont {M.~G.}\ \bibnamefont
  {Clerc}}, \ and\ \bibinfo {author} {\bibfnamefont {S.}~\bibnamefont
  {Barbay}},\ }\href {\doibase 10.1109/jstqe.2019.2929187} {\bibfield
  {journal} {\bibinfo  {journal} {IEEE J. Sel. Topics Quantum Electron.}\
  }\textbf {\bibinfo {volume} {26}},\ \bibinfo {pages} {1} (\bibinfo {year}
  {2020})}\BibitemShut {NoStop}%
\bibitem [{\citenamefont {{Robertson}}\ \emph {et~al.}(2020)\citenamefont
  {{Robertson}}, \citenamefont {{Wade}}, \citenamefont {{Kopp}}, \citenamefont
  {{Bueno}},\ and\ \citenamefont {{Hurtado}}}]{RobertsonIJSTQE20}%
  \BibitemOpen
  \bibfield  {author} {\bibinfo {author} {\bibfnamefont {J.}~\bibnamefont
  {{Robertson}}}, \bibinfo {author} {\bibfnamefont {E.}~\bibnamefont {{Wade}}},
  \bibinfo {author} {\bibfnamefont {Y.}~\bibnamefont {{Kopp}}}, \bibinfo
  {author} {\bibfnamefont {J.}~\bibnamefont {{Bueno}}}, \ and\ \bibinfo
  {author} {\bibfnamefont {A.}~\bibnamefont {{Hurtado}}},\ }\href@noop {}
  {\bibfield  {journal} {\bibinfo  {journal} {IEEE J. Sel. Topics Quantum
  Electron.}\ }\textbf {\bibinfo {volume} {26}},\ \bibinfo {pages} {1}
  (\bibinfo {year} {2020})}\BibitemShut {NoStop}%
\bibitem [{\citenamefont {der Sande}\ \emph {et~al.}(2017)\citenamefont {der
  Sande}, \citenamefont {Brunner},\ and\ \citenamefont {Soriano}}]{SandeN17}%
  \BibitemOpen
  \bibfield  {author} {\bibinfo {author} {\bibfnamefont {G.~V.}\ \bibnamefont
  {der Sande}}, \bibinfo {author} {\bibfnamefont {D.}~\bibnamefont {Brunner}},
  \ and\ \bibinfo {author} {\bibfnamefont {M.~C.}\ \bibnamefont {Soriano}},\
  }\href {\doibase 10.1515/nanoph-2016-0132} {\bibfield  {journal} {\bibinfo
  {journal} {Nanophotonics}\ }\textbf {\bibinfo {volume} {6}},\ \bibinfo
  {pages} {561} (\bibinfo {year} {2017})}\BibitemShut {NoStop}%
\end{thebibliography}%
\end{document}